\documentclass[12pt,letter]{article}
\usepackage{fullpage}
\usepackage{times}
\usepackage{epsfig}
\usepackage{latexsym}
\title{\begin{flushright}
    {UK/01-14}  \\
   {Edinburgh/2001/20}\\
{\large Dec. 2001} \\
\end{flushright}
\bigskip
The Kentucky Noisy Monte Carlo Algorithm for Wilson Dynamical Fermions}
\author{B.~Jo\'o$^a$\footnote{Current Address: Department of Physics, Columbia University, 538 W120th St,
New York, NY 10027, USA}, \ I.~Horv\'ath$^b$, and \ K.~F.~Liu$^b$\\
	\\
	$^a$Department of Physics and Astronomy \\
	University of Edinburgh, \\
	Edinburgh, EH9 3JZ, Scotland, UK. \\
	\\
	$^b$Department of Physics and Astronomy \\
	University of Kentucky\\
	Lexington, KY 40506, USA}
\begin{document}
\maketitle
\begin{abstract}
We develop an implementation for a recently proposed Noisy Monte Carlo 
approach to the simulation of lattice QCD with dynamical fermions 
by incorporating the full fermion determinant directly. Our algorithm 
uses a quenched gauge field update with a shifted gauge coupling 
to minimize fluctuations in the trace log of the Wilson Dirac matrix. 
The details of tuning the gauge coupling shift as well as results for 
the distribution of noisy estimators in our implementation are given. 
We present data for some basic observables from the noisy method, as 
well as acceptance rate information and discuss potential autocorrelation 
and sign violation effects. Both the results and the efficiency of 
the algorithm are compared against those of  Hybrid Monte Carlo.  \\
%
\end{abstract}

\newpage
\newcommand{\DU}{{\mathcal{D}U}}
\newcommand{\Dphid}{{\mathcal{D}\phi^{\dagger}}}
\newcommand{\Dphi}{{\mathcal{D}\phi}}
\newcommand{\Deta}{{\mathcal{D}\eta}}
\newcommand{\Obs}{{\mathcal{O}}}
\newcommand{\sgn}{{\rm sgn}}
\newcommand{\Pacc}{P_{\rm acc}}
\newcommand{\Pc}{P_{\rm c}}
\newcommand{\Peq}{P_{\rm eq}}
\newcommand{\Nf}{N_{\rm f}}
\newcommand{\Est}[1]{{E\left[ #1 \right]}}
\newcommand{\Wils}[1]{{\mathcal{W}_{\Box}\left(#1\right)}}
\newcommand{\Tr}{{\rm Tr}}
\newcommand{\Nrho}{N_{\rho}}
\newcommand{\Up}{U_{\Box}}
\section{Introduction}\label{s:Introduction}

Monte Carlo (MC) calculations in lattice QCD with dynamical fermions
are notoriously time consuming. These simulations generally proceed
through a numerical realization of an ergodic Markov process having the
the desired lattice QCD probability distribution as its fixed
point. In direct approaches, the major stumbling block is the
evaluation of the fermion determinant which is typically needed somewhere
in the process. For interesting volumes $V$, the fermion matrix is
extremely high dimensional and the time to compute the determinant
exactly scales as $V^3$. Hence computing the fermion determinant
exactly is not a feasible option.

The current standard workhorse for dynamical lattice QCD computations is the
Hybrid Monte Carlo (HMC) algorithm~\cite{AdkHMC}. In this case
the problem of evaluating the fermion determinant is sidestepped
by expressing the determinant as an integral over bosonic (pseudo--fermion)
fields which become full-fledged dynamical fields in the Markov
process. One criticism of the HMC method is its supposed inability
to deal with an odd number of fermion flavors. Indeed, the natural
settings for HMC are even-flavor theories  where the pseudo--fermion
heatbath is straightforward and the bosonized action is manifestly
positive. However, this limitation is not fundamental and can be
addressed within the framework of molecular dynamics
algorithms~\cite{RHMC,OddHMC}. This is a topic of current research.

Even though the direct simulation of the fermion determinant is
infamous for being nearly impossible to implement, it promises
distinct advantages over the pseudo--fermion method employed in HMC.
Besides being able to accommodate any number of flavors, it has the
potential of being a viable finite density algorithm in the canonical
ensemble approach. The usual finite chemical potential algorithm
in the grand canonical ensemble has the well-known sign problem
and the imaginary chemical potential approach has the overlap
problem~\cite{Alford}. Considering the canonical ensemble instead,
one can project out a definite baryon number from the fermion
determinant before the accept/reject step to stay in a given baryon
number sector so that the overlap problem can be avoided~\cite{LiuFD}.
In this case, it is essential to have an algorithm which accommodates
the determinant directly.

An interesting proposal for simulating the determinant directly has been
put forward recently in Ref.~\cite{FermiLab}. In that approach the idea was
to split the determinant into infrared and ultraviolet parts and treat
the infrared part exactly and the ultraviolet part approximately. This can
in principle be turned into an exact algorithm~\cite{SSDA},
but it is not yet clear how well the systematic error of the splitting 
of the determinant was under control, particularly for small quark masses and
large lattices.

The approach that will be followed here has several roots. One important
ingredient is an efficient evaluation of the determinant based on
Pad\'{e}-Z$_2$ stochastic estimators of the trace of of logarithm of the fermion
matrix~\cite{PadeZ2}. For example, using the unbiased subtraction, one can
reduce the error on the trace log of the Wilson fermion matrix on
$8^3 \times 12$ lattice at $\beta = 5.6$ by a factor of~25-40 relative
to unsubtracted one with negligible overhead. At $\kappa = 0.154$
with 400 $Z_2$ noise vectors, the absolute error of the trace log M is about
$0.29$, which translates into the same relative error for the determinant.
%

Nevertheless, this would still not be good enough if one intended to
develop a Metropolis~\cite{Metrop} like algorithm, because the
acceptance probability has to be evaluated exactly. To address this
problem, Kennedy and Kuti (KK) proposed an algorithm in which the
nonlinear Metropolis acceptance step was replaced with a linear
one~\cite{KennedyKuti}. This opened up the possibility of using
unbiased noisy estimators for the required probability ratios instead
of having to evaluate them exactly. Indeed, the required unbiased
estimators can be developed based on the idea of stochastic series
summation~\cite{BhanotKennedy}. However, the quantity used as the KK
linear acceptance probability can in principle be negative or greater
than one~\footnote{Naturally, when this is the case the quantity in
question fails to be a probability. We refer to the problem of the
acceptance probabilities being negative or greater than one as {\em
low and high probability bound violations} respectively.} when the
noisy estimate comes from the outlying tails of the underlying
distribution. This introduces a bias in the results but the
authors of Refs.~\cite{KennedyKuti,BhanotKennedy} argue that in
practice it is possible to tune both the expression for the linear
acceptance probability and the estimators so that the bias is
substantially smaller than the statistical errors.

The above discussion motivates the second root of our approach which
amounts to choosing stochastic variables so that they provide
unbiased estimators for the determinant itself (rather than acceptance
probability), which eliminates the need for the linear acceptance
step, and allows these variables to be treated as full-fledged
fields in the Markov process. This has been accomplished in
Ref.~\cite{LiuNMC} and resulted in a procedure without probability bound violation problems. We will refer to algorithms
based on this approach as Kentucky Noisy Monte Carlo (KNMC) algorithms.
In Ref.~\cite{LiuNMC} this idea was applied to a simple five state model
where the amount of noise in the estimators could be precisely tuned.
Although the statistical errors in the results of the KNMC method grew with
increasing levels of noise, the result did remain unbiased while the bias
in the KK procedure was substantially greater than the KNMC errors.

Applying this approach to QCD requires not only a satisfactory way of
estimating the determinant, but also an efficient way of proposing
new configurations in the Markov process. 
Indeed, one can easily construct a useless algorithm when proposed configurations are almost always rejected. 
It is well known that changes of the gauge field
constructed from sweeps guided solely by a pure gauge action can lead to a widely
fluctuating determinant and an essentially vanishing acceptance probability
for small quark masses. To address this issue we adopt the idea of
splitting the short-distance part of the determinant by the loop
action and incorporating it into the pure gauge
action~\cite{SSDA, HasDeg,Weingarten}. This is the third ingredient
of our approach. As a matter of fact, one of our points is that while we have
only split the determinant with the simplest plaquette action, we
nevertheless obtain a working algorithm at least for heavier quark masses.
We view the inclusion of  optimized higher loop actions for the split as
being the most promising way of improving our algorithm further.

In what follows, we present the results from applying the KNMC algorithm
with the above specifics to two flavors of Wilson dynamical fermions.
Even though the number of flavors is a mere parameter in our approach,
we use the two-flavor setting to be able to compare to HMC easily.
The remainder of this paper is organized as follows. We begin by outlining
the main ideas on which our algorithm is built in section \ref{s:Algorithm}.
We then discuss the concrete application of the algorithm to Wilson
fermions in section \ref{s:QCDApplication} where we discuss some of the
numerical techniques used in our implementation as well as some work
estimates. After presenting some computational details in section
\ref{s:CompDetails} we discuss our assorted numerical results in sections
\ref{s:RefHMC}, \ref{s:StochExpStudy} and \ref{s:NMCStudy}.
Finally we discuss these results in sections \ref{s:Discussion} and
\ref{s:UnaddressedIssues} and present our conclusions in section \ref{s:Conclusions}.

\section{The Algorithm}\label{s:Algorithm}
We start by describing the basic ideas on which our algorithm is
built.  Our goal is to simulate a distribution given by the gauge
invariant function of lattice link variables of the form
\begin{equation}
  P^{\rm QCD}(U) \,\propto\, e^{-S_g(U)} \, \prod_{f=1}^{N_{f}} \det M_{f}(U) \,=\,
                        e^{-S_g(U) + \sum_{f=1}^{N_{f}} \ \Tr \ln M_{f}(U)}
\end{equation}
where $S_g(U)$ is the gauge action and $M_{f}(U)$ is the fermion matrix ($\det
M_{f}(U) > 0$)~\footnote{In our general discussion we assume that determinant is
positive for arbitrary gauge background. For Wilson fermions (unlike overlap 
fermions) this is not strictly satisfied but the rare encounter of negative sign 
can be in principle be monitored and taken care of by including the sign into 
the observables. In our case this standard strategy will be adopted for other 
reasons anyway.} for a given flavor of dynamical quark.
The indices $f$ run over the number of flavors one
wishes to simulate. For clarity of the discussion and notation below,
we shall describe our algorithm using just a single flavor of fermion
and drop the subscript $f$ for now, with the understanding that the
generalization to many flavors is straightforward.

It will be assumed that there is a suitable approximation $R_M(U)$ of
$\ln M(U)$ that is easy to evaluate, and whose accuracy can be
controlled so that the corresponding distribution
\begin{equation}
  P(U) \,\propto\, e^{-S_g(U) + \Tr \ R_M(U)}
\label{eq:5}
\end{equation}
is arbitrarily close to $P^{\rm QCD}(U)$.

We will construct an exact algorithm for $P(U)$ of Eq. (\ref{eq:5}) based on the following considerations:

\noindent {\bf 1.} 
As pointed out in the introduction, the exact computation of $\Tr \ R_{M}(U)$
is not feasible. For this reason one would like
to use noisy estimators of this quantity. Let us consider
\begin{equation}
x \equiv E[ \Tr \ R_{M}(U), \eta ] = \eta^{\dagger} R_{M}(U) \eta \label{e:xdef}
\end{equation}
where $\eta$ is a vector in the linear space of $M(U)$ whose elements are random numbers drawn
%
%
from a distribution $P^{\rm \eta}(\eta)$ satisfying the property that
\begin{equation}
\langle \eta^{\dagger}_{i} \eta_{j} \rangle_{P^{\rm \eta}(\eta)} = \delta_{ij} \label{e:deltafn} \ .
\end{equation}

In equation (\ref{e:deltafn}) the subscripts on the angle brackets
imply that the expectation value is to be taken in the measure defined by
$P^{\eta}(\eta)$. In equation (\ref{e:xdef}) the
notation $E[ \Tr \ R_{M}(U), \eta]$ is also introduced, which may be used throughout this paper
to indicate that a given quantity is an unbiased estimator for the
first argument in the square brackets depending on the subsequent
arguments. In this case, for example, $x$ is an estimator of $\Tr \ R_{M}(U)$,
depending on the noise vector $\eta$.

From equations (\ref{e:xdef}) and (\ref{e:deltafn}) it is
straightforward to show that $x$ is indeed an estimator for
$\Tr \ R_{M}(U)$. However for simulating the measure defined by equation
(\ref{eq:5}) estimates of the quantity
$e^{x}$ are needed.  If a sequence of estimates $x_{i} =
\eta_{i}^{\dagger} R_{M}(U) \eta_{i} \ , $ with $i = 1, 2 \ldots \ $, for $x= \Tr \ R_M(U)$  is available to
us, where the subscripts on $\eta$ now refer to the position of
$\eta_i$ in the sequence rather than its elements, we can construct an
estimator for $e^{\Tr \ R_M(U)}$ by evaluating the function
\cite{BhanotKennedy}
\begin{eqnarray}
f(U, \{ \eta_i \}, \{ \rho_k \}, c) =\, 1 + \left\{ x_1
  + \theta \left(\frac{c}{2} - \rho_2\right) \left\{ {{x_2}\over {c}}
  + \theta \left(\frac{1}{3}c - \rho_3\right) \left\{ {{x_3}\over {c}} + \ldots \right. \right. \right. \nonumber \\ 
 \left. \left. \left. + \ldots \theta \left(\frac{c}{n} -\rho_{n} \right) \left\{{{x_n}\over {c}} + \ldots \right\} \right\} \right\}  \right\}
\label{e:stochExp}
\end{eqnarray}
where $c > 0$ is a tunable constant, $\theta$ is the Heavyside
step function and the $\rho_{k}$ are random number uniformly distributed
in the range $0 \le \rho_{k} \le 1$ (in other words, $\rho_k$ has distribution $P^{\rho}(\rho_k) = \theta(\rho_k) - \theta(\rho_k-1)$ \ .)  One can easily verify that
\begin{equation}
\langle f(U, \{ \eta_{i} \}, \{ \rho_{k} \},c) \rangle_{\prod_{i=1}^{\infty}P^{\eta}(\eta_i) \prod_{k=2}^{\infty}P^{\rho}(\rho_k)} = e^{\Tr \ R_{M}(U)} \ .
\end{equation}

\noindent {\bf 2.} Motivated by the discussion above, and by 
the form of equation (\ref{eq:5}), we 
extend the variable space and write the corresponding partition function 
in the form
\begin{equation}
   Z = \int dU \, e^{-S_g(U)}
       \prod_{i=1}^\infty d \eta_{i} \ P^{\rm \eta}(\eta_{i})
       \prod_{k=2}^\infty d \rho_{k} \ P^{\rm \rho}(\rho_{k})\;
       f(U,\eta,\rho)  \ ,
   \label{eq:10}
\end{equation}
where we have introduced the shorthand $f(U, \eta, \rho)$ for $f(U, \{ \eta_{i} \}, \{ \rho_k \}, c)$.
We have thus introduced an infinite number of auxiliary variables. How
can one deal with them in a practical simulation? The point is that given
the nature of terms in Eq. (\ref{e:stochExp}) only a finite number of
them will be used in any particular evaluation of $f(U,\eta,\rho)$, 
since the series terminates stochastically. 
The average number of terms can be tuned by appropriate choice of
the constant $c$, and if the typical values of $x_k$ can be kept 
reasonably small during the simulation then
a practical scheme with effectively finite number of noise fields present 
can be developed.

\noindent {\bf 3.} The basic problem with partition function (\ref{eq:10})
is that $f(U,\eta,\rho)$ is not positive definite, causing the well-known
difficulties to standard simulation techniques. We will assume (and 
demonstrate later) that things can be arranged so that the occurrence
of negative $f(U,\eta,\rho)$ in typical equilibrium configurations
$(U,\eta,\rho)$ is very small. In that case one can cure this problem by
absorbing the sign into the observables in the usual way, i.e.
\begin{equation} \label{eq:6}
\langle \Obs \rangle_P  \;=\; 
\frac{\;\langle\, \Obs\, \sgn(P)\,\rangle_{|P|}\;}
     {\;\langle\, \sgn(P) \,\rangle_{|P|}\;}   \ .
\end{equation}
Our goal then is to find a suitable Markov process for generating the 
probability distribution
\begin{equation}
   P(U,\eta,\rho) \;\propto\;
       e^{-S_g(U)} \, | f(U,\eta,\rho) |  \, 
       \prod_{i=1}^\infty P^{\rm \eta}(\eta_i) \,
       \prod_{k=2}^\infty P^{\rm \rho}(\rho_k) \ .
   \label{eq:15}
\end{equation}

\noindent {\bf 4.} One may attempt to simulate the distribution (\ref{eq:15})
in several possible ways. To explain the approach adopted here,
let us introduce the collective notation $\xi \equiv (\eta,\rho)$, \
$f(U, \xi) \equiv f(U, \eta, \rho)$, \ $P(U,\xi) \equiv P(U, \eta, \rho)$.
We can then write schematically $P(U,\xi) \propto P_1(U) P_2(U,\xi) P_3(\xi)$ with
\begin{equation} \label{e:P1P2P3Def}
P_1(U)  \propto e^{-S_g(U)} \qquad
P_2(U, \xi) \propto  | f(U, \xi) | \qquad
P_3(\xi) \propto \prod_{i=1}^\infty P^{\eta}(\eta_{i})\,
                 \prod_{k=2}^\infty P^{\rho}(\rho_k)  \ . 
\end{equation}
We will use two steps based on the following two statements that can be 
verified directly:

(a) Let $T_1(U,U')$ be the ergodic Markov matrix satisfying
detailed balance with respect to $P_1$, in other words
$P_1(U) T_1(U,U') dU = P_1(U')T_1(U',U) dU'$. Then the transition matrix
\begin{equation} \label{e:MetropGauge}
    T_{12}(U,U') \;=\; T_1(U,U')\,  
    \min\, \Bigl[\, 1,{{P_2(U',\xi)}\over{P_2(U,\xi)}}\,\Bigr]
\end{equation}
satisfies detailed balance with respect to the $P_1(U) P_2(U,\xi)$ 
(with $\xi$ fixed).

(b) The transition matrix
\begin{equation} \label{e:MetropNoise}
    T_{23}(\xi,\xi') \;=\; P_3(\xi')\, 
    \min\, \Bigl[\, 1,{{P_2(U,\xi')}\over{P_2(U,\xi)}}\,\Bigr]
\end{equation}
satisfies detailed balance with respect to $P_2(U,\xi) P_3(\xi)$
(with $U$ fixed).

From (a), (b) it follows that $T_{12}$ and $T_{23}$ keep the original 
distribution $P(U,\xi)$ invariant and interleaving them will lead to
an ergodic Markov process with the desired fixed point.

We note that there is a lot of freedom in choosing the pure gauge
process $T_1(U,U')$. If local updates are used, then it is necessary
to ensure that a given sequence of such updates satisfies detailed
balance with respect to $P_1(U)$. This can be achieved for example by
updating the sites at random or selecting the order of updated
variables appropriately. We adopt the procedure wherein only links
corresponding to chosen even/odd part of the lattice and chosen
direction are updated.  One can easily check that such a
``sub--sweep'' satisfies detailed balance for the Wilson pure gauge
action if the elementary local updates also do so.  Further, we note
that in step (b) use is made of the fact that the probability
distribution $P_3(\xi)$ for the noise can be generated directly from a
heatbath.

Finally it should be emphasized that in equations (\ref{e:MetropGauge}) and 
(\ref{e:MetropNoise}) one needs to compute a ratio of the form:
\begin{equation}
\frac{P_2(U', \xi)}{P_2(U, \xi)} = \frac{| f(U', \xi)|}{ | f(U, \xi) |}
\end{equation}
where $f(U, \xi)$ in Eq. (\ref{e:stochExp}) is an estimator for $e^{x}$. Since the quantity
$x$ is an estimator for the quantity $\Tr \ R_M(U)$,
it can be very large, as $R_M(U)$ is an extensive quantity. Looking at 
equation (\ref{e:stochExp}) it can be seen that $f(U,\eta, \rho)$ can indeed
give a very poor estimate of the exponential, if the $x_{k}$ are 
large, and only a few terms are taken. 

Ideally one would like to be in a situation where $-1 < x_{k} < O(1)$. Certainly when $x_{k} < -1$, one faces the
problem that $f(U, \eta, \rho)$ can become negative depending 
on the number of terms taken. If this happens only
occasionally the effects can be taken into account by folding the sign
of $f(U, \eta, \rho)$ into the observable as in equation
(\ref{eq:6}). However, if it happens often, it can cause a large
effective reduction in statistics.

While no firm upper limit has been placed on $x_{k}$ we do note that 
the exponential function diverges rapidly for increasing $x > 0$. 
Given an infinite amount of statistics, the stochastic exponentiation 
technique will still give an unbiased estimator for $e^{x}$. However,
when $x > 1$ the terms in (\ref{e:stochExp}) have increasing absolute value, 
thus causing the variance of the estimators to become very large. Furthermore, 
in a Markov process such as the one described above, the evolution can potentially get 
stuck in a region of configuration space with a given number of terms 
(noise fields $\rho$) being used to estimate $f(U, \eta, \rho)$. This is because 
although having a large number of terms is unlikely, once reached with $x_{k} > 1$, 
then $f(U, \eta, \rho)$ will have a higher numerical value than it would with fewer 
terms (corresponding to a potential new noise field configuration $\rho'$) in which 
case the new field is likely to be rejected. For this reason it is prudent in 
a simulation to arrange matters so that $x_{k}$ is of $O(1)$. 

The above discussion suggests that, while the approach described above
theoretically leads to simulating the distribution (\ref{eq:5}),
additional steps need to be taken to turn it into a practical scheme.
We now discuss some ways that can be employed to deal with the issue of typical
magnitudes and variances of $x_n$ below.

\subsection{Shifting the Action by a constant}
Motivated by the fact that a ratio of exponentials can be written as
\begin{equation}
\frac{e^{x'}}{e^{x}} = \frac{e^{(x' - x_0)}}{e^{(x - x_0)}} \ , 
\end{equation}
one notices that the fermionic action can be shifted by a constant
through making the replacement:
\begin{equation}
x(U, \eta) = \eta^{\dagger} R_{M}(U) \eta \ \rightarrow \ x(U, \eta, x_0) = \eta^{\dagger} R_{M}(U) \eta - x_0 .
\end{equation}
Such a shift can move the mean of the distribution of the values of
$x$ to an arbitrary real number without affecting the simulation in
any way. With this in mind, our main goal is to minimize the variance
of $x$.

\subsection{Splitting with the Loop Action}
It is well known that a significant portion of $\Tr \ln M(U)$ can be
typically taken into account by a short-distance loop action $\Delta S_{g}(U)
$ ~\cite{HasDeg,Weingarten}, especially at larger quark masses, and this
is expected to remain true for $\Tr \ R_M(U)$ also.  This fact
can be used to reduce the magnitude of the fluctuations in $x$ by splitting
this part of $\Tr R_M(U)$ into the gauge action when setting up the
Markov process (see e.g.~\cite{SSDA}). To recall the argument let us write
\begin{equation}
  P(U) \,\propto\, e^{-S_g(U)}\, e^{\Tr \ R_M(U)} \,=\,
         e^{-S_g(U) + \Delta S_g(U)}\,
         e^{\Tr \ R_M(U) - \Delta S_g(U)}
\end{equation}
We can thus replace
\begin{equation}
   S_g(U)     \;\longrightarrow\;  S_g(U) - \Delta S_g(U)   \qquad\qquad
   \Tr \ R_M(U) \;\longrightarrow\;  \Tr \ R_M(U) - \Delta S_g(U)  
\end{equation}
in our Monte Carlo procedure. Then the gauge updates are performed with the
new local action, and evaluation of $f(U,\eta,\rho)$ involves the variables
$x_n$ estimating $\Tr \ R_M(U) - \Delta S_g(U)$. The specifics
of how to do this will be discussed in section \ref{s:LoopSplittingImplementation}

\subsection{Explicit Splitting}
Utilizing the fact that $e^x = (e^{x/N})^N$, one can also split $\Tr \
R_M(U)$ directly by writing $\Tr \ R_M(U) = \sum_{i=1}^N \frac{1}{N} \Tr \  R_M(U)$,
and use separate noise fields for every $\frac{1}{N} \Tr \ R_M(U)$.  Since
$N$ divides $\Tr \ R_M(U)$ into $N$ pieces, each carrying $1/N$ flavor, we
shall refer to it as the number of fractional flavors.  Indeed, the
corresponding modification of Markov process is straightforward. To
see this, consider for simplicity the case $N=2$. Originally, the
simulated probability distribution was written schematically as
$P(U,\xi) \propto P_1(U) P_2(U,\xi) P_3(\xi)$, while now we have
\begin{displaymath}
  P(U,\xi_1,\xi_2) \propto
  P_1(U) P_2^s(U,\xi_1) P_2^s(U,\xi_2) P_3(\xi_1) P_3(\xi_2),
\end{displaymath}
where the $P_2^s$ is $P_2$ of equation (\ref{e:P1P2P3Def}) with $x$ 
from Eq. (\ref{e:stochExp}) replaced by $x/N$.

In the step (a) of the MC procedure  we thus have
$P_2 \longrightarrow P_2^s(U,\xi_1) P_2^s(U,\xi_2)$ with $\xi_1,\xi_2$
fixed. There is an arbitrariness in selecting the process (b). For
example if one chooses to update a single set of noise at a time,
step (b) does not change at all, and one can choose for example the 
sequence (a), (b)$_{\xi_1}$, (a), (b)$_{\xi_2}$ as an elementary 
Markov step. The only requirement here is the overall ergodicity.

The main effect of explicit splitting is to scale the width of the
distribution of $x$ by the number of fractional flavors $N$. 
The optimal mixture of action shifting, splitting by the loop action
and explicit splitting is a matter to be explored.

\subsection{Reducing the Variance from Noise}
While splitting the loop action as described above reduces the
fluctuations in $x$ arising from the fluctuation of the gauge
configurations, the variance of $x$ also receives a contribution
from the noise fields $\eta$ since $\eta^{\dagger}R_M(U)\eta$ is used
in the construction of $x$.  Further
variance reduction techniques can be applied to reduce this contribution.
The particular technique depends on the kind of noise used. In 
the specific case when $Z_2$ noise is used, it has been shown \cite{bmt94}
that all the contributions to the variance of $\Tr \ R_{M}(U)$ come
from off diagonal elements of $R_M(U)$ in which case the
unbiased subtraction noise reduction technique of \cite{PadeZ2}
is highly effective. We will present details of this method in 
section \ref{s:EstimatingRM}.

\section{Application to Lattice QCD with Dynamical Wilson Fermions} \label{s:QCDApplication}

To demonstrate that the ideas described in the previous section can
lead to a working algorithm, we now describe the details of the 
implementation of the algorithm that we used to perform 
simulations with two flavors of degenerate Wilson quarks. Although
in principle both the algorithm and the implementation 
can handle an arbitrary number of flavors, the case of two degenerate flavors
is convenient from the point of view that its results can be checked 
against HMC simulations. Further, we can also carry out some tuning
using these reference simulations as we shall detail in sections \ref{s:LoopSplittingImplementation} and \ref{s:RefHMC}.
 
We simulate the theory with the standard Wilson gauge action
\begin{equation}
S_{g}(U) = -\frac{\beta}{3} {\rm Re} \ \Tr \ \Up
\end{equation}
where $\beta$ is the gauge coupling parameter. The quantity $\Up$ is obtained 
as usual by evaluating the product of link matrices around each elementary plaquette 
and summing the results over the whole lattice. After integrating out the Grassmann 
numbers, the effective fermion action is
\begin{equation}
S_{f}(U) = - \sum_{f=1}^{N_f} \Tr \ \ln \ M(U, \kappa_{f})
\end{equation}
where the sum is over all desired flavors, $M(U, \kappa_{f})$ is
the Wilson fermion matrix
\begin{equation}
M(U, \kappa_f) = 1 - \kappa_f D(U) \ ,
\end{equation}
$D(U)$ is the usual Wilson hopping matrix and $\kappa_f$ is
the hopping parameter for the flavor with index $f$. In our
simulations we used an approximation $R_{M}(U)$ to $\ln M(U)$ 
given by a Pad\'e approximation, which we will discuss in more detail
in section \ref{s:EstimatingRM}.

\subsection{Local Gauge Update}
In order to update the gauge fields, we use the quasi heatbath method \cite{QuenchedHeatbathAlgorithm}
amended as described previously. We choose one particular parity (even
or odd) of our lattice and one of the 4 space--time dimensions at
random, and update all the links with that parity and in that direction
simultaneously. Each such sub--sweep allows us to update $\frac{1}{8}$
of our lattice. As outlined earlier, one is  free to perform any number
of such updates before updating the noise fields. In fact, this remains
a free parameter ($N_{s}$) in our code. However for the results presented
here we have always used $N_{s} = 1$.

\subsection{Estimating $\Tr \ R_M(U)$} \label{s:EstimatingRM}
In order to estimate $\Tr \ R_M(U, \kappa)$ we turn to the 
technology described in \cite{PadeZ2}. The logarithm is approximated 
using a Pad\'e approximation, which after a partial fraction expansion, 
has the form:
\begin{equation} \label{e:Pade}
\ln M(U, \kappa) \approx R_{M}(U) \equiv  b_0 \ I - \sum_{i=1}^{N_{P}} b_{i}\left(M(U, \kappa) + c_{i} \ I \right)^{-1}
\end{equation}
where $N_{P}$ is the order of the Pad\'e approximation, and the constants 
$b_{i}$ and $c_{i}$ are the Pad\'e coefficients. In our implementation 
we have used an 11-th order approximation whose coefficients are 
tabulated in \cite{PadeZ2}.

The traces are then estimated by evaluating bilinears of the form $\eta^{\dagger} R_{M}(U) \eta$. If the
components of $\eta$ are chosen from the $Z_2$
group,  then the contributions to the variance of these bilinears come only from
off diagonal elements of $R_M(U)$ as discussed previously.
In this case\footnote{In this sense $Z_2$
noise is optimal. With other types of noise such as Gaussian noise,
the variance receives contributions from diagonal terms which one
cannot subtract off. In this case the unbiased subtraction scheme
described here is ineffective.} an effective method reducing the variance is to subtract off a linear combination
of traceless operators from $R_{M}(U)$ and to consider
\begin{equation}
E[ \Tr \ R_{M}(U), \eta ] = \eta^{\dagger} \left( R_{M}(U) - \omega_{i} \Obs_{i} \right) \eta  \ .
\end{equation}
 Here the $\Obs_{i}$ are
operators with $\Tr \ \Obs_{i} = 0$.  Clearly since the $\Obs_{i}$ are
traceless they do not bias our estimators in any way. The $\omega_{i}$
are constants, that can be tuned {\em a priori} to minimize the fluctuations
in $E[ \Tr \ R_{M}(U), \eta]$.

In practice the $\Obs_{i}$ are constructed by taking traceless terms
from the hopping parameter expansion for $M^{-1}(U)$. These reduce the
noise coming from the terms $\left( M(U) + c_{i} \right)^{-1}$ in equation
(\ref{e:Pade}). The terms $D$, $D^{2}$, $D^{3}$ and further odd
powers of $D$ are explicitly traceless and terms which have even
powers such as $D^{4}$ have known traces given in terms of various
loops. For example
\begin{equation}
\Tr  \ D^{4}(U) = -64 \Tr \ \Up
\end{equation}
and hence $\Obs_{4} = D^{4}(U) + 64 \ \Tr \ \Up$ is traceless.
Details of finding the traces of even powers of $D$ can be found, for
example, in \cite{MontvayMunster}.
In our computations we have subtracted observables involving 
$D$, $D^2$, $D^3$, $D^4$, $D^{5}$ and $D^{7}$.

Although the parameters $\omega_{i}$ are tunable in principle, the 
hopping parameter expansion for $M^{-1}(U)$ is sufficiently good 
for heavier quark masses, that numerically the $\omega_{i}$ are close
to unity. Hence in our simulations we have always used $\omega_{i} = 1$
for all $i$.

Since in our implementation we need the sum of $R_M(U)$ for all
flavors, we can estimate the whole sum using a single noise field
$\eta$. This allows us to compute all the $(M(U, \kappa_f) + c_i
I)^{-1} \eta$ for all $c_{i}$ and all flavors $\kappa_f$, for a given
$\eta$ using a single multiple shift\footnote{Also known as
multi--mass.} inversion \cite{Jegerlehner}. In practice we employ the
$M^{3}R$ \cite{Jegerlehner, Glassner} algorithm as it is the most
memory efficient, and memory was a bottleneck issue on our target
QCDSP architecture.

\subsection{Loop Splitting Specifics} \label{s:LoopSplittingImplementation}
We now turn to the details of splitting the loop action. 
The fermionic action for a single flavor can be written as
\begin{equation}
S_{f}  = -\left( \Tr \ R_M(U, \kappa_f) - \lambda^f {\rm Re} \ \Tr \ \Up \right) - \lambda^f {\rm Re} \ \Tr \ \Up ,
\end{equation}
where the $\lambda^{f}$ is a tunable parameters for that particular flavor.
One can then shift the fermion action for each flavor as follows:
\begin{equation}
S_{f}(U) \rightarrow - \left( \Tr \ R_M(U, \kappa_f) - \lambda^{f}\ {\rm Re} \ \Tr \ \Up \right) \ .
\end{equation}
At this point it becomes convenient to introduce the shorthand $T(U, \lambda^f)$
for the quantity
\begin{equation} \label{e:TDefn}
T(U, \lambda^f) \equiv \Tr \ R_M(U, \kappa_f) - \lambda^f \ {\rm Re} \  \Tr \ \Up \ 
\end{equation}
and to write
\begin{equation}
S_{f}(U) \rightarrow - T(U, \lambda^f) \ .
\end{equation}

In order to absorb this change, the gauge action needs to be
correspondingly shifted as
\begin{equation}
S_{g}(U) \rightarrow -\frac{\beta}{3} {\rm Re} \ \Tr \ \Up - \lambda^f \ {\rm Re} \ \Tr \ \Up = - \frac{(\beta + 3\lambda^f)}{3} \ {\rm Re} \ \Tr \ \Up
\end{equation}
with an extra shifted term for each flavor of fermion. The end result
is that the gauge action becomes:
\begin{equation}
S_{g}(U) = -\frac{\beta'}{3} {\rm Re} \Tr \ \Up \quad \mbox{with} \quad \beta'=\beta + 3 \sum_{f=1}^{N_f} \lambda^{f} \ .
\end{equation}

The $\lambda^{f}$ need to be tuned to minimize the variance of
$T(U,\lambda^{f})$.  The tuning procedure is given by the action
matching technology of Sexton, Irving and Weingarten
\cite{Weingarten,SextonIrving}. In fact, finding $\lambda_{{\rm
min}}^f$, the values of $\lambda^f$ for which the fluctuations of
$T(U,
\lambda^{f})$ are minimized, corresponds exactly to tuning a quenched
simulation to a dynamical fermion one in an action matching sense. The
quantity $\lambda^{f}_{\rm min}$ is given (see \cite{SextonIrving}) by
the formula
\begin{equation} \label{e:SextonIrvingTuning}
\lambda^{f}_{{\rm min}} = - \frac{{\rm Cov}(\Tr \ R_M(U, \kappa_f), {\rm Re} \ \Tr \ \Up )}{\sigma^{2}({\rm Re} \ \Tr \ \Up)}
\end{equation}
where $\sigma^2({\rm Re} \ \Tr \ \Up)$ is the variance of the plaquette
and the quantity in the numerator is the standard covariance between
the $\Tr \ R_M(U, \kappa_f)$ and the plaquette.

We note, the action matching technology of \cite{SextonIrving}, is not
limited to simply tuning the Wilson plaquette action, but is fairly
generic.  In particular, it can be used to tune the splitting of the
determinant by actions which are linear combinations of higher order
loops.

When a preliminary reference simulation is available at the desired
parameters, one can measure the required covariances and correlations
on this data--set. Otherwise, since the tuning of \cite{SextonIrving}
can be carried out in any measure, one can perform a quenched simulation,
and employ a self consistent procedure to find $\lambda^{f}_{\rm min}$. 

Once $\lambda^{f}_{\rm min}$ are determined, one can immediately compute
$\langle T(U, \lambda^{f}_{\rm min}) \rangle$
which are good first estimates for the action shift parameters $x^{f}_{0}$, 
which will ensure the quantities
$x^{f} = E[ T(U, \lambda^{f}_{\rm min}) ] - x^{f}_0$
have means of $0$. These may not be the optimal shift factors $x^{f}_0$, since
it may be desirable to have $\langle x^{f} \rangle > 0$, to minimize number
of negative sign violations.

One can then shift the  $x^{f}$ even  further so that practically all
the values of $x^{f}$ are greater than 0.
This can be achieved by defining
\begin{equation}
x^{f}_{0} = \langle T(U, \lambda^{f}_{\rm min}) \rangle - \frac{1}{N_{f}}x_{1}\ . \label{e:xDefn}
\end{equation}
where $x_{1}$ is some factor of $\sigma(E[ T(U,\lambda^{f}_{\rm min})] )$.

The final value $x$ that we use in equation (\ref{e:stochExp}) is then
\begin{equation} \label{e:defxreally}
x = \frac{1}{N} \sum_{f} E[ T(U, \lambda^{f}) ] -  x^{f}_0 \ 
\end{equation}
with $x^{f}_0$ as defined in equation (\ref{e:xDefn}).

For later reference, the values and definitions of all the parameters in
our implementation are summarized in Table \ref{t:Params}.
\begin{table}
\begin{center}
\leavevmode 
\begin{tabular}{|cl|}
\hline \hline 
Parameter & Description \\ 
\hline
$\beta$     & Gauge Coupling \\
\hline
$\kappa_{f}$ & Fermion Hopping parameter (1 per flavor) \\
\hline
$N_{\eta}$ & Number of noise vectors per estimator of$ E[ R_{M}(U) ]$ \\ 
           & (we use $N_{\eta}=1$) \\
\hline 
$\omega_{i}$ & Parameters for the reducing the noise in \\ 
             & $E[ \Tr \ R_M(U), \eta ]$\\
	     & (we use $\omega_{i} = 1$ for all $i$) \\
\hline
$ r  $       & Target fractional residual in the multiple mass inverter \\
	     & (we use $r = 10^{-6}$ for the lightest shifted mass) \\
\hline
$ \lambda^{\rm min}_{f}$ & Loop action splitting parameters \\
			 & ( 1 per flavor ). The shifted gauge coupling \\
	                 & is $\beta' = \beta + 3 \sum_{f} \lambda^{f}_{\rm min}$ \\
\hline
$  N  $ & Number of fractional flavors (explicit splitting terms) \\
\hline
$ x^{f}_0 $ & Action shifting constants   \\ 
            & (1 per flavor) \\
\hline
$x_1$    & Action shift fine tuning factor \\
	 & (we use $x_1 = 2$)\\
\hline
$ c $    & Variance control parameter for equation (\ref{e:stochExp}) \\
	 & (we use $c = 1.5$) \\
\hline
$ N_{s}$ & The number of scalar sub--sweep 'hits' \\
         & in the gauge update algorithm \\
	 & (we used $N_{s} = 1$) \\
\hline
\end{tabular}
\end{center}
\caption{Summary of implementation parameters}
\label{t:Params}
\end{table}

\subsection{Work Estimates}
The cost $C$ of the present implementation of KNMC for each accepted
update of the gauge field and noise fields is
\begin{equation} \label{cost}
C \sim \frac{N_{\eta}\, N\, N_{exp} C_{\rm M} + C_{\rm G}}{P_{\rm acc}^U} + \frac{N_{\eta} \, N_{\rm exp} C_{\rm M}}{P_{\rm acc}^{\xi}} \ .
\end{equation}
In Eq. (\ref{cost}) above, the first term represents the computational
cost of updating the gauge field, and the second corresponds to the 
contribution from updating a single noise field (out of the $N$).
Here $N_{\rm exp}$ is the average number of terms in the stochastic
expansion of the exponential function in Eq. (\ref{e:stochExp}) which
is $e$ for the case $c = 1$.  $C_{\rm M}$ is the cost of estimating
$\Tr \ R_{M}(U)$ for all flavors but for only one noise field, $C_{\rm G}$ is the cost of updating the gauge
configuration $U$. The quantities $P_{\rm
acc}^U$ and $P_{\rm acc}^{\xi}$ are the acceptance rates for the gauge
and noise updates respectively.  The cost $C_G$ is negligible in
comparison with $C_{\rm M}$ which is dominated by the time to perform the 
multiple mass solution of the system $( M(\kappa) + c_{i} ) X = \eta$ for 
all $\kappa$ and $c_{i}$.

\subsection{Volume Scaling}

The cost for creating a single estimator for $x$ is dominated by the
cost of the multiple mass solve. This should scale linearly with the
volume. The quantity $x$ itself is expected to scale with the square
root of the volume, since evaluating the bilinear involves a sum of
random numbers over the volume which can be positive or negative with
an equal likelihood. Hence one would expect the variance $\sigma^2(x)$
of $x$ to scale linearly with $V$ and so $\sigma(x)$ should scale as
$\sqrt{V}$. In this case the number of fractional flavors needed to
keep $\sigma(x)$ to be $O(1)$ must also increase as $\sqrt{V}$. Hence
the total cost of the algorithm must scale at least as $O(V^{3/2})$.

\subsection{Comparison to HMC} 
Let us compare our work estimate to that of a typical HMC accepted  
configuration. The work involved for grows as
\begin{equation} \label{e:ApproxHMCCost}
C_{\rm HMC} = \frac{N_{\rm MD} C_{\rm F} + 2 C_{\rm H}}{P_{\rm acc}^U}
\end{equation}
where $N_{\rm MD}$ is the number of time--steps one takes while
integrating the Hamiltonian equations of motion for one trajectory.
The predominant contribution to the cost of such a time--step is 
the computation of the molecular dynamics force for the time--
step, which for fermionic systems involves solving the system
of equations: $(M^{\dagger}M) x = \phi$,  where $\phi$ are the pseudo--fermion
fields. The cost $C_{\rm H}$ is the cost of calculating the energy
which also requires the solution of a system similar to that of the
force computation. The energy calculations are done at the start and end of the trajectory.
While in principle one can carry out the inversions for the energy
using a different stopping criterion from the one used for the
force computation, it is convenient now to consider a case
where this is not done and $C_{\rm F}=C_{\rm H}$. 

Since the predominant cost for our KNMC algorithm (assuming that
$P_{\rm acc}^{U} \le P_{\rm acc}^{\xi}$) comes from the accept/reject
step following the gauge field update; when the determinant has to be
estimated for all the fractional flavors (c.f. the first term of
Eq. (\ref{cost})\ ) we will neglect the cost of updating a single
noise field (where the determinant only has to be estimated for a
single fractional flavor -- c.f. the second term of
Eq. (\ref{cost}). Also, as $C_{\rm G}$, the cost of performing the
gauge update sweep, is negligible in the current implementation in
comparison to $C_{\rm M}$, the cost of performing a multiple mass
inversion, the cost of the noisy algorithm is approximately
\begin{equation} \label{e:ApproxNoisyCost}
C_{\rm KNMC} \sim \frac{N_{\rm \eta} \, N \, N_{\rm exp} C_{\rm M}}{P_{\rm acc}^{U}} \ .
\end{equation}

Comparing Eqs. (\ref{e:ApproxNoisyCost}) and (\ref{e:ApproxHMCCost})
and assuming that $C_{\rm M} \sim C_{\rm F}$ since they both involve
a solution of a similar set linear equations we note that
the two algorithms are comparable when
\begin{equation}
\frac{N_{\eta}\, N \, N_{\rm exp}}{P^{\rm KNMC}_{\rm acc}} \sim \frac{N_{\rm MD}}{P^{\rm HMC}_{\rm acc}}
\end{equation}
where $P^{\rm KNMC}_{\rm acc}$ refers to the gauge acceptance rate of the
KNMC algorithm and $P^{\rm HMC}_{\rm acc}$ refers to the HMC acceptance
rate. In a typical application, $N_{\rm MD} \sim O(100)$ and $P^{\rm HMC}_{\rm acc} \sim 0.8$. As we shall see later on, our simulations using the 
KNMC algorithm managed to achieve $P^{\rm KNMC}_{\rm acc} \sim 0.3$, with
$N_{\rm exp} \sim 3$ and $N \approx 20$, which makes our current 
simulations somewhat more expensive than their HMC counterparts. 

\section{Computational Details} \label{s:CompDetails}
We now briefly describe our numerical computations. In all
we have performed 5 numerical studies. One of these was a reference 
HMC simulation, another was a brief study of the behavior of the
stochastic exponentiation technique and the remaining three were 
KNMC computations.
 
Our implementation of the KNMC algorithm was coded for the QCDSP
\cite{QCDSP} supercomputer, and was run on 1, 2 and 4 motherboard
QCDSP computers located at Columbia University and the Brookhaven National
Laboratory. Our code was written in C++ utilizing the Columbia Physics
Software System which was made available to us by the RIKEN--BNL--Columbia 
(RBC) Collaboration.

Our reference Hybrid Monte Carlo (HMC) simulation (see below) was carried out
at the T3E facility at NERSC, using the GHMC \cite{UKQCDHMC} code made available to us by the UKQCD collaboration. 

Our analysis program, as well as our investigation
of the stochastic exponentiation was carried out on workstations.

\section{Reference HMC Simulation} \label{s:RefHMC}
In order to carry out the required tuning, and to have some benchmark
results for our noisy simulations we have performed a reference
HMC simulation with two flavors of Wilson dynamical fermions using the
desired physical simulation parameters listed in Table \ref{t:HMCParams}. 
We generated 1280 HMC trajectories, of which the first 625 were discarded
for equilibration. Of the remaining 655 trajectories we stored every fifth 
one to measure $\Tr \ R_M(U)$ giving us a total of 132 configurations 
to work with.
On these configurations we have estimated $\Tr \ R_M(U)$ using 100
noise vectors per configuration. When the noise fields per configuration
were averaged, the measurement of $\langle \Tr \ R_M(U) \rangle_{\eta}$
was accurate to a relative error of less than 1\% per configuration.

\subsection{HMC Observables}

\begin{table} 
\begin{center}
\begin{tabular}{|cc|}
\hline \hline Parameter & Value  \\
\hline
$\beta$  & $5.5$ \\
$\kappa$ & $0.1550$ \\
$N_{f}$ & 2 \\
$V$ & $8^4$ \\
\hline
\end{tabular}
\end{center}
\caption{Physical Parameters for our Simulations}
\label{t:HMCParams}
\end{table}

\begin{table} 
\begin{center}
\begin{tabular}{|cc|} 
\hline  \hline Observable & Value  \\
\hline
$\langle {\rm Plaquette} \rangle$ & $0.5476(1)(4)$ \\
$\langle \Tr \ R_M(U) \rangle$ & $640.9(6)(20)$ \\
\hline
\end{tabular}
\end{center}
\caption{Reference HMC Results. The firs set of errors are the naive
bootstrap errors. The second set shows the effects of autocorrelation
estimated by blocking the data.}
\label{t:HMCResultsObs}
\end{table}

The values of $\Tr \ R_M(U)$ and the plaquette (normalized by the
volume and the number of planes) measured in our HMC computations are
shown in Table \ref{t:HMCResultsObs}.  In the case of the plaquette,
we used the values of the observable on all $655$ trajectories. The
statistical errors were first estimated using a simple bootstrap
technique with $500$ bootstrap samples.  A blocking technique was then
used to estimate the effects of autocorrelation on the observables.
This technique consisted of averaging successive values of the
observable in the time series into a single observable of a new
data--set (with less statistics than the original). The naive variance
was then measured on the resulting new data--set. This procedure was
repeated until we ran out of statistics, or observed a plateaus in the
variance.  Unfortunately, this data is rather noisy and hence
estimating the plateaus is somewhat subjective. We believe we have
been conservative in Table \ref{t:HMCResultsObs}.

\subsection{Tuning $\lambda^{f}_{\rm min}$}
We now describe the results of performing the tuning for the $\lambda^{f}_{\rm min}$. Since, both the HMC and the KNMC simulations were done
using degenerate flavors of fermions, we will drop the flavor index $f$ 
on this quantity from now on.

\begin{figure}
\begin{center}
\leavevmode
\hbox{
\epsfxsize=5in
\epsffile{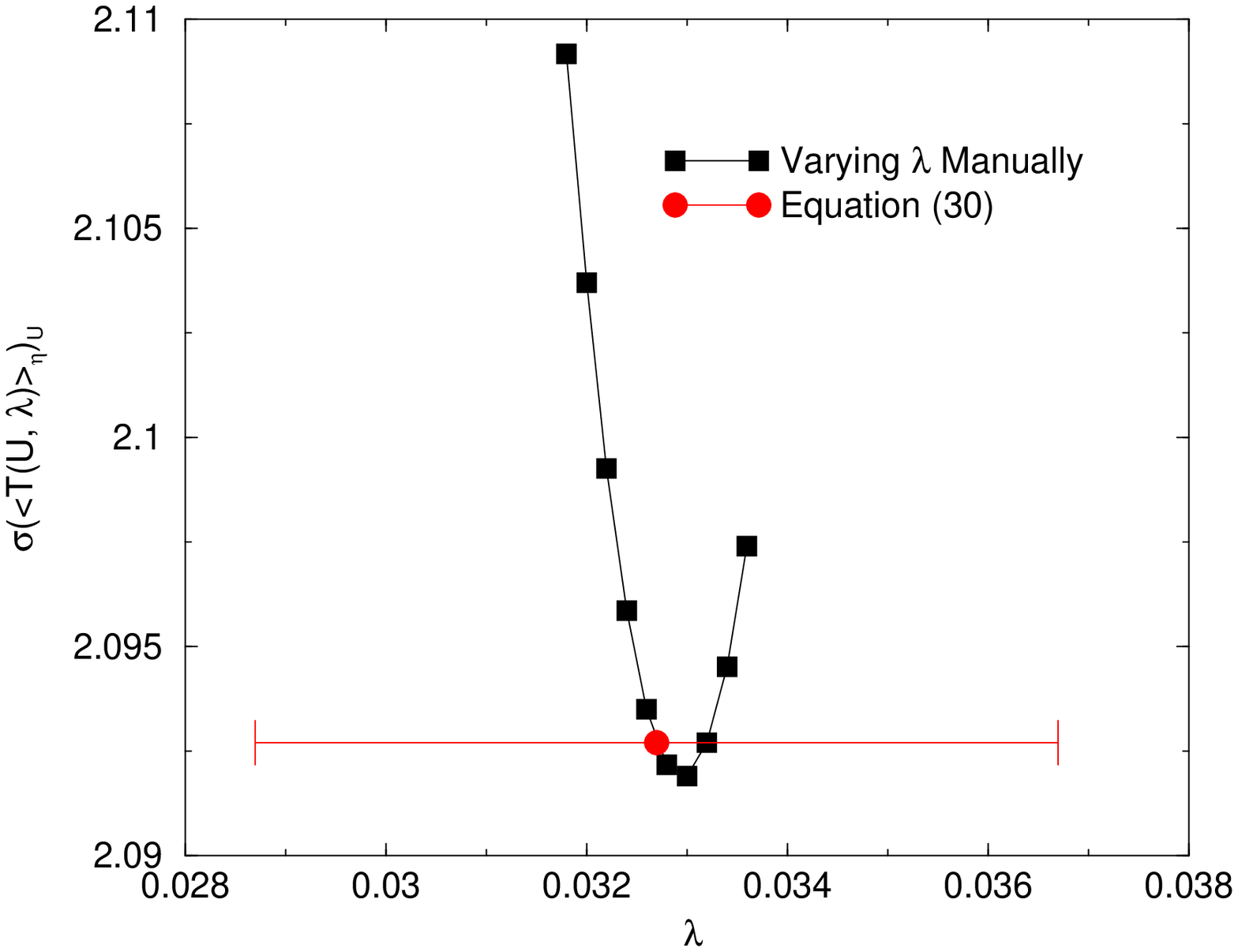}
}
\end{center}
\caption{Tuning for $\lambda_{\rm min}$ using Method 1. The circle gives
the result from equation (\ref{e:SextonIrvingTuning}). The squares are the
results of explicitly varying $\lambda$ around this minimum.}
\label{f:TuningPlotMethod1}
\end{figure}

\begin{figure}
\begin{center}
\leavevmode
\hbox{
\epsfxsize=5in
\epsffile{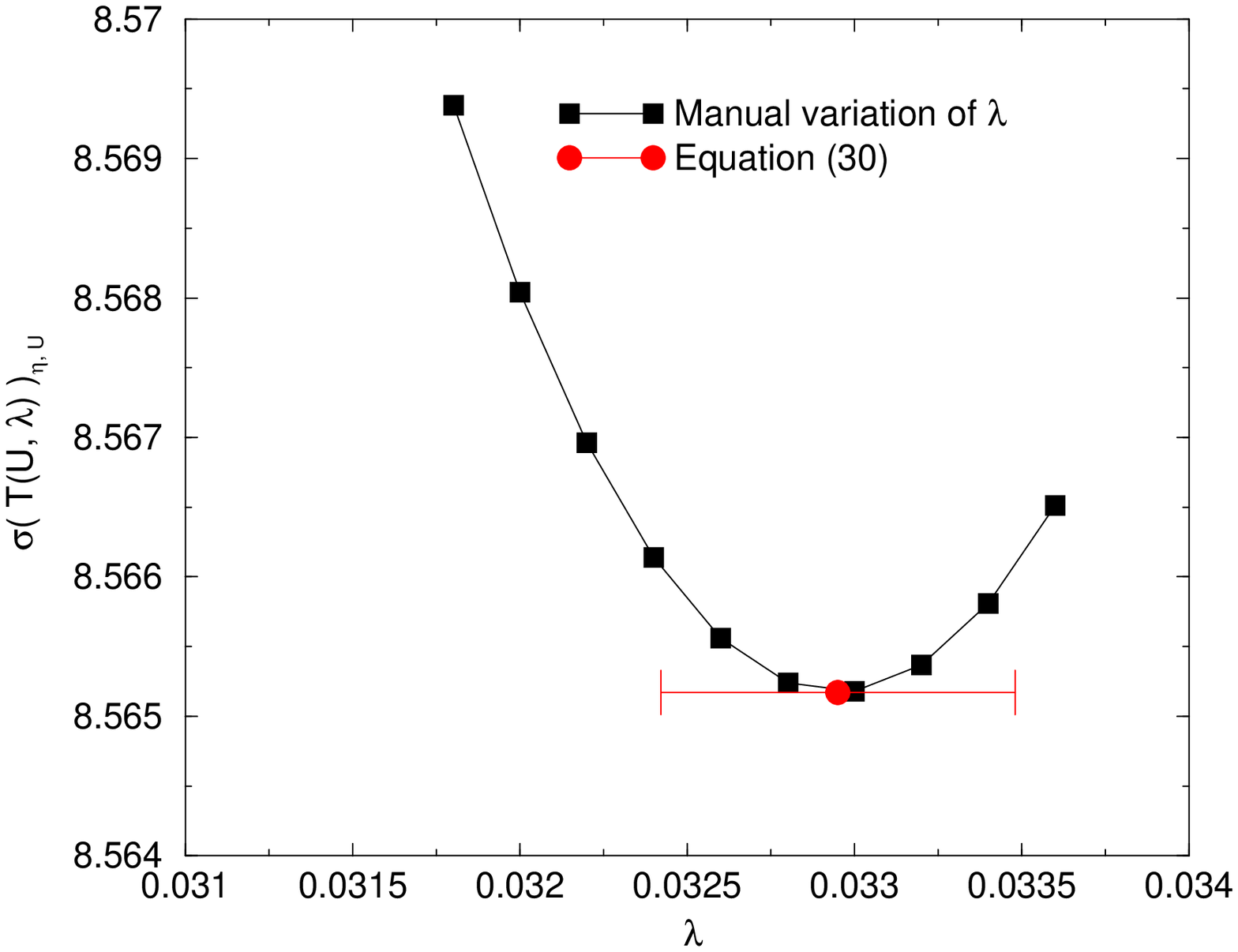}
}
\end{center}
\caption{Tuning for $\lambda_{\rm min}$ using Method 2. The circle gives
the result from equation (\ref{e:SextonIrvingTuning}). The squares are the
results of explicitly varying $\lambda$ around this minimum.}
\label{f:TuningPlotMethod2}
\end{figure}

We used the estimators of $\Tr \ R_M(U)$ to estimate $\lambda_{\rm
min}$ using the tuning formula of equation
(\ref{e:SextonIrvingTuning}). Before outlining the results we note
that there are two ways of computing the variances and covariances in
equation (\ref{e:SextonIrvingTuning}), the choice of which has a
bearing on the resulting standard deviation, $\sigma(T(U, \lambda_{\rm
min}^f))$ of $T(U, \lambda_{\rm min}^f)$:
\begin{enumerate}
\item{{\bf Method 1: \ }} In this method, all the estimators $E[ \Tr \ R_M(U) , \eta]$ are first averaged over all the noise fields $\eta$
for a given configuration. This gives an estimate of $\langle \Tr
\ R_M \rangle_{\eta}$ per configuration with some small error. These estimates can then be used (neglecting the small errors) to 
perform averages with respect to the gauge fields as usual when
computing variances, covariances and correlations.  The results for
$\sigma(T(U, \lambda))$ as a function of $\lambda$ for this method are
plotted in Fig. \ref{f:TuningPlotMethod1}.

\item{{\bf Method 2: \ }} In this method one does not first average over the $\eta$
fields. Instead the averaging is performed over all the noise
fields and gauge fields simultaneously when evaluating variances,
covariances and correlations. The results for $\sigma(T(U, \lambda))$
are plotted in Fig. \ref{f:TuningPlotMethod2} for this method.
\end{enumerate}

While Method 1 is perhaps the preferred method from the point of view
of action and observable matching, the numbers from it may be misleading
from the point of view of a noisy algorithm since it neglects the effects
of noise in the estimation of $\Tr \ R_M(U)$.  However, one would expect the 
two methods to both give the same $\lambda_{\rm min}$ as in effect 
they are both equivalent to carrying out the same path integral. In 
Method 2 since more statistics are available, one may 
expect to get smaller errors on $\lambda_{\rm min}$. Finally comparing
the results of methods 1 and 2, one can get a rough idea of how 
much of the variance in our $\Tr \ R_M(U)$ comes from the noise fields $\eta$
and how much comes from fluctuations from gauge configuration to gauge configuration.

We note in passing, that methods 1 and 2 can be thought of 
as opposite extremes of carrying out KNMC simulations with various
values of $N_{\eta}$. Method 1 corresponds to the situation where 
$N_{\eta}$ is large, and many conventional estimators $E[ \Tr \ R_M(U); \eta ]$ are averaged, to get a better estimator, whereas method 2 corresponds
to the situation where $N_{\eta}=1$.

Looking at Table \ref{t:HMCTuningRes} it can be seen that the two
methods do in fact give similar results for $\lambda_{\rm
min}$. Method 2 appears more accurate, presumably because of the
larger number of estimators available. By examining
Figs. \ref{f:TuningPlotMethod1} and \ref{f:TuningPlotMethod2} the
increase in statistics is clearly visible from the size of the
horizontal error bar on the tuned point. It can also be seen that the
minima are quite shallow in terms  of $\lambda$.  The error bar on the
point obtained with method 1 is quite large, despite the fact that the
point itself lies near the minimum. With method 2 the error bar is
smaller and the point is better placed. Our recommendation from these
results would be to always check that the minimum is found, by
performing some manual tuning around the value of $\lambda_{\rm min}$
given by the Eq. (\ref{e:SextonIrvingTuning}).

We note that we carried out the measurements of Method 2 after our
KNMC simulations as an afterthought.  Hence our simulations all used
values determined by Method 1.

\begin{table}
\begin{center}
\begin{tabular}{|l|cc|} 
\hline \hline     Statistic      & Method 1          & Method 2 \\ 
\hline
$\sigma(\Tr \ R_M(U))$              & $7.99$            & $11.51$ \\
Corr$(\Tr \ R_M(U), \Tr \ \Up)$   & $0.96(12)$	     & $0.66(1)$ \\
\hline
$\lambda_{\rm min}$ \ ($\times 10^{-2}$) & $3.27(38)$ & $3.29(5)$ \\ 
\hline
$\sigma(T(U,\lambda_{\rm min}))$ & $2.09$ & $8.56$ \\
 $\langle T(U, \lambda_{\rm min})\rangle $ & $-679.6(1)$ & $-689.60(7)$ \\
\hline
\end{tabular}
\end{center}
\caption{HMC Tuning Results}
\label{t:HMCTuningRes}
\end{table}

\section{Stochastic Exponentiation Study} \label{s:StochExpStudy}
Before we describe our KNMC simulation results, we will make another
detour and experiment with the technique of stochastic
exponentiation. A question of interest is: How good an estimator $E[
e^{x} ]$ of $e^{x}$ can one obtain by applying equation
(\ref{e:stochExp}) to estimators $E[x]$ of $x$.  In this section we
attempt to give a partial answer to this question in a situation where
both $x$ and its fluctuations, as characterised by its standard deviation
$\sigma(x)$, are under explicit control.

In this study, noisy estimates $E[x]$ were made for several
values of $x$ by adding Gaussian noise of known variance $\sigma^2(x)$
to the actual values of $x$. Eq. (\ref{e:stochExp}) was then applied to these
values of $E[x]$ to make estimators $E[ e^{x}]$ of $e^{x}$. 

\begin{figure}
\begin{center}
\leavevmode
\hbox{
\epsfxsize=5.5in
\epsffile{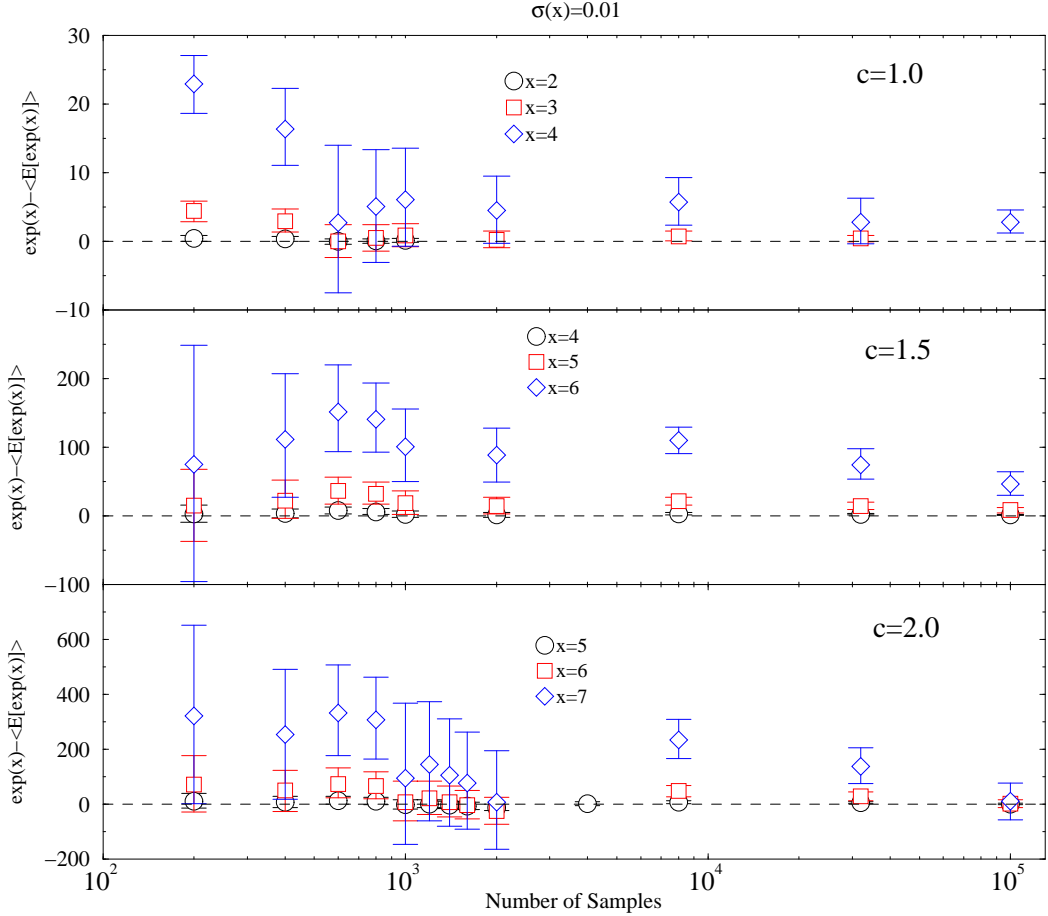}
}
\end{center}
\caption{Bias in the Stochastic Estimation as a function of statistics}
\label{f:Bias}
\end{figure}

The results of this study are shown in Fig. \ref{f:Bias} where the
bias in the results of the stochastic exponentiation is plotted
against the number of samples of $E[e^{x}]$.  To be more precise, a number of
samples of $E[e^{x}]$ were averaged to obtain a measurement of
$\langle e^{x} \rangle$ and this was subtracted from the true value of
$e^{x}$. It can be seen from Fig. \ref{f:Bias} that the technique
works quite well for $x=3$, $c=1$ and about 1000 samples.  Increasing
$c$ to $c=1.5$ allows one to get unbiased estimates for $x=5$ for the
same number of samples and it is even possible to get unbiased estimates for
$x=6$ for such a sample size if $c=2.0$. However we note that as $x$
is increased the fluctuations increase enormously too, as can be seen
when one compares the scales on the vertical axes of
Fig. \ref{f:Bias}.  The data shown in Fig. \ref{f:Bias} confirm our
earlier reasoning about the distribution of $x$ in our earlier
discussions, namely that it is preferable for the value $x$ in
Eq. (\ref{e:stochExp}) to be small.

\section{KNMC Simulations} \label{s:NMCStudy}
We now turn to the discussion of our KNMC simulations. In all, 
three major simulations have been carried out, to which we shall refer
as $S1$, $S2$ and $S3$ respectively. In all three simulations we 
have used the same physical parameters as in our trial HMC simulation
(see Table \ref{t:HMCParams}). We used the loop splitting factor $\lambda_{\rm min} = 3.27 \times 10^{-2}$ for both flavors  as obtained by Method 1 of the tuning (see Table \ref{t:HMCTuningRes}). For our value of $x^{f}_{0}$,
we used $\langle T(U, \lambda_{\rm min}) \rangle = -679.6$ (Table \ref{t:HMCTuningRes}) with an additional
fine tuning factor of $x_1 = 2$ as per equation
(\ref{e:xDefn}).

Using this value of $\lambda_{\rm min}$, resulted in a gauge coupling
shift of $\Delta \beta = 3 N_{f} \times \lambda^{f}_{\rm min} = 0.1962$ giving
a value of $\beta' = 5.6962$ to use in the quenched gauge updating algorithm
(instead of the $\beta = 5.5$ of the HMC computations).

The only difference between the three KNMC simulations was the value of the
number of fractional flavors $N$ which took the values $N=15$, $N=20$
and $N=25$ for simulations $S1$, $S2$ and $S3$ respectively. This
choice was based on the values of $\sigma(T(U,\lambda_{\rm min}))$ measured in
the preliminary HMC simulation using Method 1.

We show some basic statistics for the simulation in Table
\ref{t:NMCParams}. In particular, we give the number of negative signs for
$f(U, \eta, \rho)$ that we counted along each simulation and the width
of the distribution of $x$ as characterised by its standard deviation
$\sigma(x)$, with $x$ defined as in Eq. (\ref{e:defxreally}).
%

\subsection{Distribution of $x$}
\begin{figure}
\begin{center}
\leavevmode
\hbox{
\epsfxsize=5.5in
\epsffile{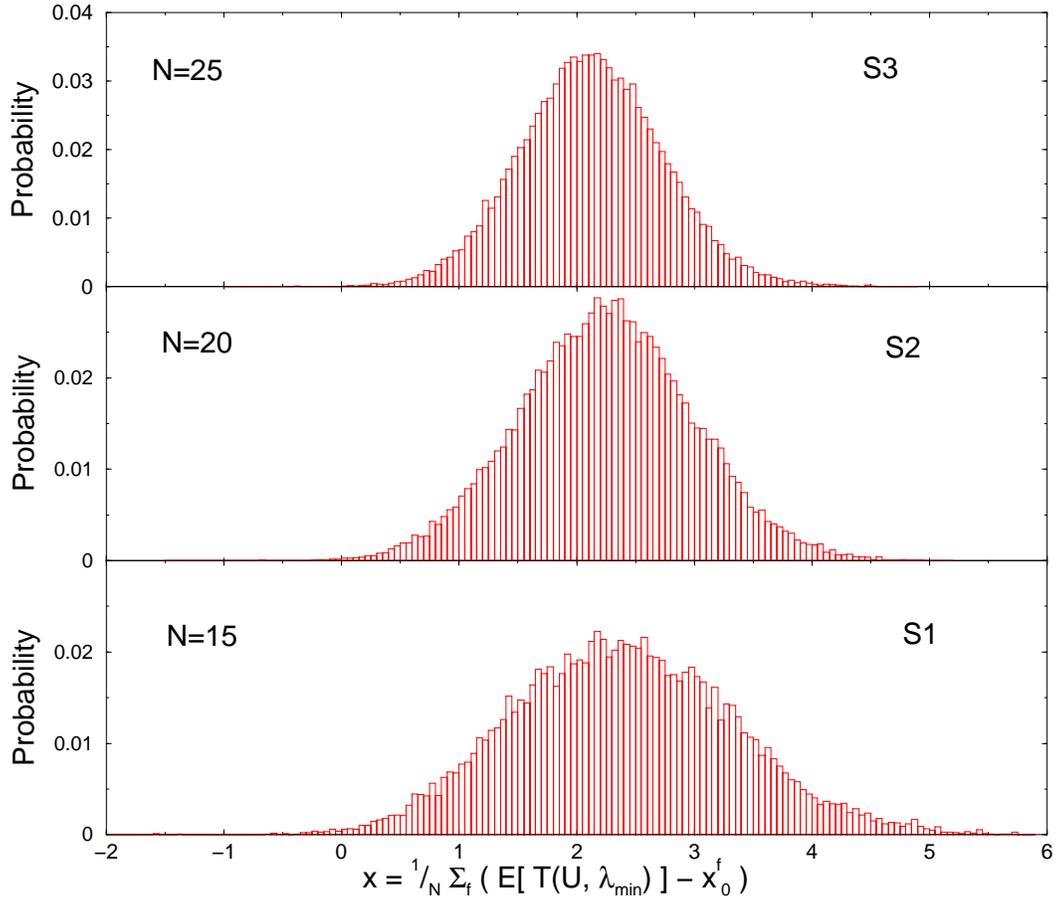}
}
\end{center}
\caption{Distributions of $x$ for the three noisy simulations}
\label{f:TrLnDist}
\end{figure}

In Fig. \ref{f:TrLnDist} we plot the distributions of the quantity $x$
as measured in the three simulations. The distributions
appear to be Gaussian as one would expect from the Central Limit Theorem.

It can clearly be seen, that simulation $S1$ is quite near the limits prescribed
upon the values of the quantity $x$ by the stochastic exponentiation study, 
namely that the values of $x$ are getting near the upper limit of $x=4$,
$x=5$ where the stochastic exponentiation technique begins to break down
for our limited statistics.
Also for simulation $S1$, it can clearly be seen that the lower tail of the distribution
stretches well beyond 0. This manifests itself in that about 
$2.9\%$ of the estimators for $E[ e^{x}]$ were negative which, has a noticeable effect on the statistical errors for observables as will be demonstrated shortly.

Simulation $S2$ seems to be more or less where one would expect this
noisy method to behave well. A few of the estimates for $x$ are larger
than $x=5$ and although the tail of the distribution
stretches into the negative region, in practice this results in very
few sign violations of $f(U, \eta, \rho)$ (Only 1 out of the total
number of statistics equating to $0.02\%$). The trick of folding
the sign of $f$ into the observable may be a practical proposition in this case.

Finally $S3$ is the best behaved of the simulations, with few values of
$x > 4$ and no sign violations in $f(U,\eta, \rho)$. The results
of this simulation can be analysed with conventional techniques.

\begin{table}
\begin{center}
\begin{tabular}{|lcccc|}
\hline \hline
Simulation & N & \# Gauge Updates &  negative signs of $f(U, \eta, \rho)$ & $\sigma(x)$ \\
           &   &                  &  ( Number, \%-age )  &  \\
\hline
   S1      & 15 & 2400 &  (70, 2.9) & 0.944 \\
   S2      & 20 & 4229 &  (1, 0.023) & 0.734 \\
   S3      & 25 & 4050 &  (0, 0)   & 0.6   \\
\hline
\end{tabular}
\end{center}
\caption{Summary of statistics for the noisy simulations}
\label{t:NMCParams}
\end{table}

\subsection{Acceptance Rates}
The acceptance rates of the three KNMC simulations are shown in
Table \ref{f:Acceptances}. One can see that the gauge acceptance
rate seems not to depend on the number of fractional flavors
used ($N$), whereas there is a marked increase in the noise update
acceptance rate when $N$ is increased. We believe that being 
able to achieve a gauge acceptance rate of around 33\% by 
performing quenched updates at a shifted $\beta$ is a great success
of the action matching technology, however, for the algorithm
to be practical it is somewhat low. Such a low acceptance rate, combined
with updating only $\frac{1}{8}$--th of the lattice gauge 
fields with every update can result in very long autocorrelation times
(as will be discussed in section \ref{s:ACs}). Clearly for the algorithm to be practicable, 
a better gauge update scheme is needed.

\begin{table}
\begin{center}
\begin{tabular}{|lccc|}
\hline \hline Simulation & $N$ & Gauge Update  & Noise Update \\
                         &   & Acceptance (\%) & Acceptance (\%) \\
\hline
	$S1$		& $15$ &  $32(1)$ & $49(1)$ \\
        $S2$   		& $20$ &  $33(1)$ & $53(1)$ \\
	$S3$		& $25$ &  $33(1)$ & $55.7(7)$ \\
\hline
\end{tabular}
\end{center}
\caption{Acceptance Rates for the noisy simulation}
\label{f:Acceptances}
\end{table}

\subsection{Observables}

\begin{figure}
\begin{center}
\leavevmode
\hbox{
\epsfxsize=5.5in
\epsffile{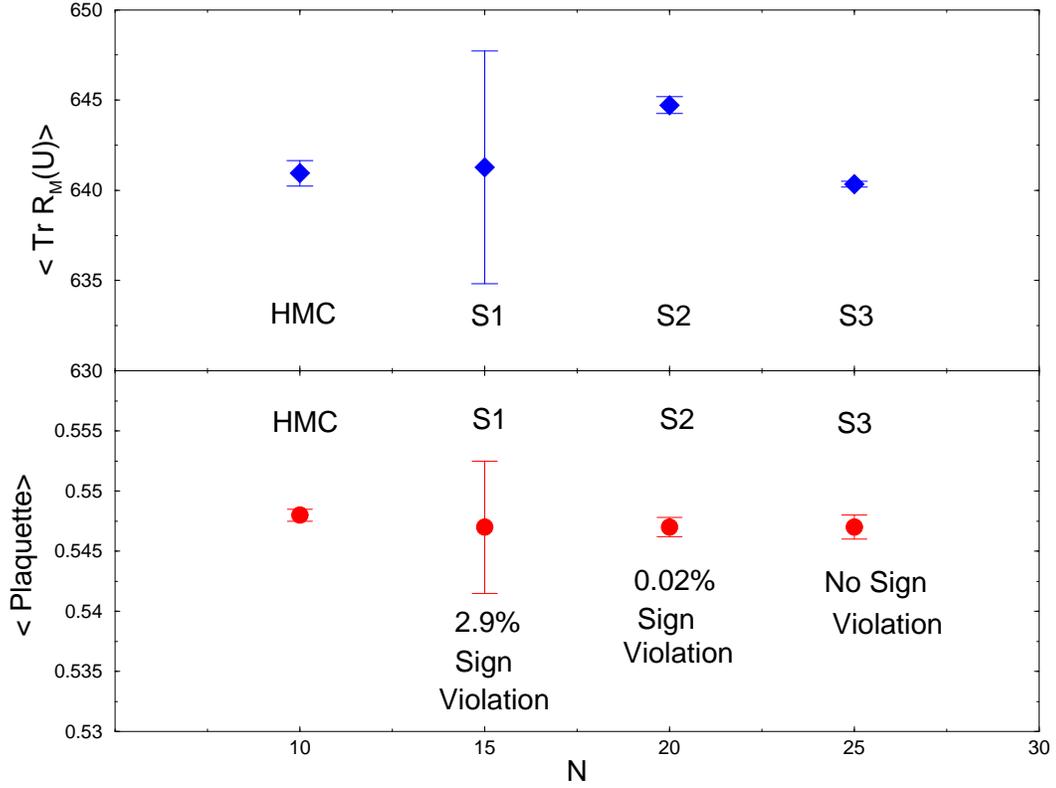}
}
\end{center}
\caption{Observables from the KNMC simulations. The plaquette is shown
on the bottom graph, and $\langle \Tr \ R_M(U) \rangle$ is shown on the 
top. The values plotted at $N=10$ are the HMC results for comparison.}
\label{f:NoisyObs}
\end{figure}

In Fig. \ref{f:NoisyObs} we show our measurements of the plaquette
and $\Tr \ R_M(U)$ for the KNMC simulations as well as the result of
the reference HMC calculation for comparison.  The error estimates for
the noisy simulations do not include the effects of autocorrelations
so as not to obscure the effects incurred by the sign violations in
$f(U,\eta, \rho)$.

We note with gratification, that the results for simulation $S1$ appear
unbiased, even with  $2.9\%$ of the estimates of $f(U, \eta, \rho)$ having
negative signs. However the errors on this value are massive when compared
to those of the other simulations.
%

In Table \ref{t:NoisyPlaqErrors} we show the bootstrap errors on the
numerator and denominator of Eq. (\ref{eq:6}) used in evaluating
the expectation value of the plaquette in the presence of sign
violations (for $S1$ since $S3$ is free of sign violations and there is
only 1 single violation in $S2$). In the third line 
we tabulate the relative error in the plaquette
measurements when the sign is not folded in -- although it must be borne
in mind that doing the analysis this way would give a biased
value for the plaquette.

In the case of simulation $S1$ it can clearly be seen that the magnitude
of the relative errors when the sign is folded in is about two orders
of magnitude greater than when it is not, and that the relative errors
in the numerator and denominator (first two lines in Table \ref{t:NoisyPlaqErrors}) are approximately the same. This clearly suggests
that the errors are entirely dominated by the error in the sign.

\begin{table}
\begin{center}
\begin{tabular}{|c|c|cc|}
\hline \hline Simulation & \%-age sign violation & Observable & Relative Bootstrap Error \\
\hline
     &     &$\langle {\rm Plaquette} \ \sgn(f) \rangle$ & $0.709 \%$ \\ 
$S1$ & 2.9 &$\langle \sgn(f) \rangle$ & $0.708\%$ \\ 
     &     &$\langle {\rm Plaquette} \rangle$ & $0.0037\%$ \\
\hline
\end{tabular}
\end{center}
\caption{The relative errors in the numerator and denominator of the
quantity needed to estimate the expectation value of the plaquette for $S1$.
The third line shows the relative error on the plaquette without folding in the sign.}
\label{t:NoisyPlaqErrors}
\end{table}

\subsection{Autocorrelations} \label{s:ACs}
Let us now turn to the question of autocorrelations. It is not entirely
clear how to best estimate autocorrelation effects in the presence of
sign violations. When a substantial amount of sign violations are
present, one would expect these to be the dominant contributors to the
statistical error in any case. However, we did attempt to make an
investigation into autocorrelation effects in simulation $S3$
where no negative signs are present in $f(U, \eta, \rho)$.

\begin{figure}
\begin{center}
\leavevmode 
\hbox{
\epsfxsize=5.5in
\epsffile{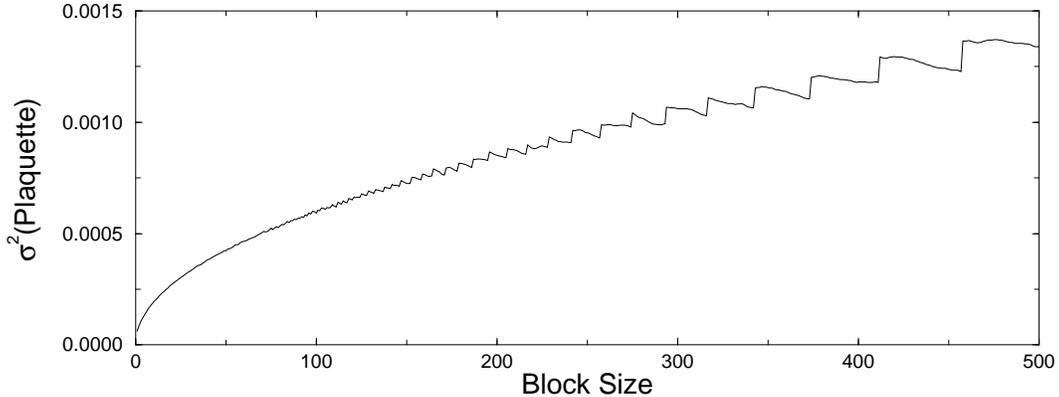}
}
\end{center}
\caption{Variance of the plaquette measurement of $S3$ as a function of block size.}
\label{f:PlaqAC}
\end{figure}

Once again, we used the blocking procedure that  was  outlined earlier
for our HMC simulation (section \ref{s:RefHMC}). The growth of the variance of the plaquette is plotted 
as a function of block size in Fig. $\ref{f:PlaqAC}$. It can be seen
that the errors do not plateau as a function of block size, indicating
that the integrated autocorrelation time is very long. We expect this is
due in part to the fact that only $\frac{1}{8}$-th of the lattice is updated
with every gauge update, and in part because the rate of acceptance for the
gauge updates is quite low -- about 33\%.

We note that in our experience with the quasi--heatbath method in
quenched simulations, the plaquette usually decorrelates in about 20-40
sweeps (at around this level of gauge coupling and on similar
volumes). However in this implementation of KNMC only
$\frac{1}{8}$--th of the lattice is updated with any one sweep.  This
in itself could be expected to increase the autocorrelation time to
about $160-320$ accepted sweeps.  This would not present a problem in
a conventional quenched simulation where gauge field updates are cheap
and no sweeps are rejected (they are after all from a heat--bath).
However when one couples a potentially expensive noisy accept/reject
after each $\frac{1}{8}$--th update the computational cost increases
significantly, so that one cannot hope to achieve the level of
statistics in quenched simulations. In this case the low acceptance
rate of the simulations becomes a problem.  Clearly for the KNMC approach
to be practicable, a better gauge update algorithm is needed  than the one 
used here, with a higher
acceptance rate.

\subsection{Checking the Tuning}
Before we proceed to summarize and discuss our numerical results, we would 
like to discuss the quality of the tuning for the three KNMC simulations. 
On physical grounds, one would expect that the parameter $\lambda_{\rm min}$
which minimizes the variance of the fermion action, is a universal quantity,
and should depend largely on the physical parameters which define the 
expectation value of $\Tr \ \ln M(U)$. The amount of variance reduction thus
achieved is expected to depend on the gauge generation algorithm to
some degree, but certainly, one would expect some self consistency
when carrying out the tuning on the HMC and the KNMC data--sets.

To this end we repeated the procedure for tuning $\lambda_{\rm min}$
using both methods 1 and 2 on estimates of $\Tr \ R_M(U)$ produced during
simulation $S3$ which is not affected by sign violations. The main
difference here is that the number of estimates of $\Tr \ R_M(U)$
varied slightly from configuration to configuration since the number
of noisy estimates for $x$ differs for each update. However, with
$N=25$ and a value of $c=1.5$ the average number of terms used in
evaluating $f(U, \eta, \rho)$ was about 65 terms per gauge update. With over
4000 updates, these statistics should prove adequate.

\begin{table}
\begin{center}
\begin{tabular}{|l|cc|} 
\hline \hline     Statistic      & Method 1          & Method 2 \\ 
\hline
$\sigma(\Tr \ R_M(U))$              & $10.38$            & $12.73$ \\
Corr$(\Tr \ R_M(U), \Tr \ \Up)$   & $0.98$	     & $0.81$ \\
\hline
$\lambda_{\rm min}$ \ ($\times 10^{-2}$) & $3.51(4)$ & $3.526(7)$ \\ 
\hline
$\sigma(T(U,\lambda_{\rm HMC}))$ & $2.0$ & $7.49$ \\
$\sigma(T(U,\lambda_{\rm min}))$ & $1.87$ & $7.45$ \\
\hline
\end{tabular}
\end{center}
\caption{KNMC Tuning Results for one flavor for simulation $S3$.}
\label{t:NMCTuningRes}
\end{table}

The re--tuning results are shown in Table \ref{t:NMCTuningRes}.  We
note that the value of $\lambda_{\rm min}$ has now increased a little
with respect to the HMC results ( Table \ref{t:HMCTuningRes}),
however, this is not a very large change.  Indeed, it is less than
10\% of the HMC value in Table \ref{t:HMCTuningRes}.

We show in Table \ref{t:NMCTuningRes} the effect of the newly
determined $\lambda_{\rm min}$ on the $\sigma$ values as well as the
$\sigma$ value from the old HMC tuning for comparison.  It can be seen
that the change in the value of $\lambda_{\rm min}$ from that of the HMC
result does not reduce the $\sigma$ value by a great deal, probably
because of the very flat minimum of $\sigma$ as a function of $\lambda$.

A similar trend can be seen, when switching to method 1 from method 2, as was
visible in the HMC case. When averaging the noise fields and effectively
measuring $\langle \Tr \ R_M(U) \rangle$ on each configuration, the subtraction of the loop
action from the plaquette is much more effective than when using method
2 -- (a reduction from $\sigma \approx 10$ to $\sigma \approx 2$ in the 
former case against a reduction from $\sigma \approx 12$ to $\sigma \approx
7.5$ in the latter).

\section{Summary of Numerical Results And Discussion} \label{s:Discussion}
\subsection{Tuning $\lambda_{\rm min}$}
Our main result here is that the tuning can be done in two ways
(methods 1 and 2) to carry out the minimization of
$\sigma(T(U,\lambda))$.  We found that a much larger degree of noise
reduction can be achieved by subtracting the loop action using method
1 than method 2. Some of the gain seen here may be spurious due to
neglecting the errors from the noise when averaging the noisy
estimators, and some of it is real and comes from the fact in method 2
 extra noise is being added to the computation, and the minimization is 
not actually carried out
with respect to $\Tr \ R_M(U)$ but with respect to noisy
estimators of it. Further, the minima thus found is very flat with
respect to $\lambda$ (see Figs. \ref{f:TuningPlotMethod1} and
\ref{f:TuningPlotMethod2}) implying that not much gain may be made by
dynamically tuning the $\lambda$ parameter.

These results seem to imply that  a greater improvement may be achieved
in the acceptance rates of the noisy algorithm using the
loop--splitting technique if more noise vectors were used in the noisy
estimators of $\Tr \ R_M(U)$ instead of the current 1 vector per
estimator (i.e. if $N_{\eta}$ was increased.) However, this would also imply
more numerical work as computing each noisy estimator involves a multi--
mass inversion. On the other hand it may be possible to reduce the number
of fractional flavors ($N$) in return. Further investigation is
required to establish when the trade--off becomes worthwhile. Finally,
it was found that switching to method 2 from method 1 on the noisy 
data--sets showed a behavior pattern very similar to switching 
between methods 1 and 2 on the HMC data, even if the actual values
were somewhat different.

\subsection{Stochastic Exponentiation Technique} 
The stochastic exponentiation technique works well when the argument
$x$ to be exponentiated is small and positive. When $x>1$ successive
terms in the expression for $f(U, \eta, \rho)$ have greater and greater
numerical value although the probability of reaching these terms still
drops factorially. This implies that the variance of the estimates is
likely to be large when $x$ is large, and also that the estimates for
the exponential are likely to be poor when only a few terms are
taken. If $x$ is negative, one risks getting negative values of
$f(U, \eta, \rho)$ which can result in large statistical errors (see
below).

\subsection{Observables and Sign Violations}
While the expectation values of observables appear to be
unbiased in or simulations, it appears that even a small number of
negative signs in $f(U, \eta, \rho)$ -- such as $2.9\%$ of the total
number of estimates -- can completely dominate the statistical
errors. In this situation the effort of creating more and more
configurations goes into reducing the error in the estimate of
$\langle \sgn(f(U, \eta, \rho) \rangle$ a more difficult problem
than the usual $\frac{1}{\sqrt{N}}$ problem of reducing the errors in the
bare observables. While in the KK linear accept/reject approach these sign
violations manifest themselves as an explicit bias in the result, in
KNMC this bias is traded for a larger statistical error.

\subsection{Autocorrelations}
Our final result is that our gauge updating algorithm performs
rather poorly. Although updating the gauge and noise fields is 
computationally easy, these updates are now coupled to a computationally
very expensive accept/reject step. The updates have long autocorrelation
times and a low acceptance. This situation needs to be seriously 
addressed if the algorithm is to be competitive with say HMC.

\section{Issues not Addressed in this study} \label{s:UnaddressedIssues}
This study was the initial foray into the study of KNMC
algorithms. There are several issues which have not been addressed which
are also relevant to the algorithm. We outline two of these here.

\subsection{Equilibration}
In our study, we have always started our simulations from an 
equilibrated configuration produced by our preliminary HMC study. 
One may very well ask the question: ``How would we equilibrate 
our algorithm and tune the necessary parameters if the reference
simulation was not present?'' We point to the idea outlined in 
\cite{SextonIrving}. The idea presented there is that one can 
carry out an initial quenched simulation, which can be used to carry
out a preliminary tuning. This will provide amongst other things a
shifted $\beta$ value. One can then carry out a second quenched
computation with the shifted $\beta$ value, thus bringing the quenched
configuration distribution as close to the intended dynamical one as
possible. At this point, one can start to carry out simulations with
the noisy algorithm, re-tuning $\beta$ and the other parameters along
the way until a self consistency is achieved. This is possible because
the tuning in \cite{SextonIrving} can be carried out in any measure.

\subsection{The question of an infinite number of noise fields}
One may be concerned that since technically  an infinite number
of dynamical noise fields are present in Eq. (\ref{e:stochExp}) it is not
possible to update them all. In particular, very high order terms in
Eq. (\ref{e:stochExp}) may never be reached. Thus some of these
noise fields will have infinitely long autocorrelation times. Another
way of saying this is that the KNMC algorithm may not be ergodic in its
infinite variable state space.  

While this is a problem in principle, we do not expect it to be a
problem in practice, since the probability of reaching the higher
order terms is factorially suppressed. Because of this suppression, we
expect that these fields can have little effect on our partition
function and that any bias in our results from such fields is expected
to be very much smaller than statistical errors. It may be possible 
to construct operators that probe these high order terms explicitly, 
where the effect of long autocorrelations should be clearly visible.
Perhaps a more relevant potential setback comes from the visibly long 
autocorrelations of our gauge update procedure. We note that in the 
KK approach, this problem does not arise, since in that case the noise 
is not part of the state space. On the other hand, algorithms adopting 
the KK accept/reject step have the in--principle problem of probability 
bound violations which can introduce a bias into the answers. Hence 
in choosing between the two approaches, one has a choice of which 
in--principle problem one wishes to accept as the challenge.

\section{Conclusions} \label{s:Conclusions}
We have developed a QCD implementation for the Kentucky Noisy Monte Carlo
approach and performed an initial numerical study in the context of two 
flavors of dynamical Wilson fermions. This study was a success in several ways, 
most notably since we have managed to assemble all the necessary numerical 
technology required for incorporating the fermion determinant directly, for the first 
time with controlled systematic errors. The method produced results that are 
consistent with reference Hybrid Monte Carlo simulations.

We have gained valuable insight into the necessary tuning methodology,
and have learned what essentially drives the algorithm, notably the
stochastic properties of the quantity $x$, which needs to be distributed 
so that it is of $O(1)$ and has a small variance. A large variance leads 
to many excessively large estimates in the tail of the distribution, 
causing the stochastic exponentiation technique to be inefficient. Also, 
on the other side of the distribution, one could get many sign violations
which, while not introducing bias, can lead to large statistical errors. 
Even though the distribution can be made arbitrarily narrow by employing
more noise fields, by using more loops for splitting the determinant, 
and by using a larger number of fractional flavors, all these come at the 
price of an increase in computational cost.

Unfortunately, in our current implementation, the algorithm is not
particularly efficient. It suffers from the problem of long
autocorrelations and rather low acceptance rates. One possible way for
addressing this issue could be the use molecular dynamics for
updating the gauge field. Once again however, this improvement would
come at a potentially high computational cost.

In addition, several other improvements have been suggested for making
the algorithm more efficient~\cite{Walter}, notably using the technique 
of eigenvalue deflation~\cite{EvalDefl} and the use of additional noisy 
techniques~\cite{NoisyInv,DeForc}, both to improve the convergence of
matrix inversions. 

Despite the above shortcomings in the current implementation (related
mainly to the efficiency), we believe that our approach holds a great
future promise with its capability to handle an arbitrary number of
fermion flavors.  Also, and perhaps more importantly, in combination with the
projection of the definite baryon number from the determinant, it is
the contending candidate for the future finite density algorithm of~\cite{LiuFD}. 

 \section{Acknowledgments}
We would like to acknowledge DOE grant DE-FG05-84ER0154 and the Center
for Computational Sciences at the University of Kentucky for financial
support. We are extremely grateful to N.~H.~Christ, R.~D.~Mawhinney,
Columbia University and the RBC
Collaboration for providing us access to the QCDSP hardware and
application code to ease our development and to provide a platform for
simulation. We would like to thank the UKQCD collaboration for
allowing us to use their GHMC code, optimized for the T3E. B.~ Jo\'o
would like to thank the Department of Physics at Columbia University
for providing support for travel between Columbia University and the
University of Kentucky under the SciDAC addition to their DOE grant
DE-FG02-92ER40699, and PPARC for financial support by way of
employment through grant PPA/G/O/1998/00621 through the later stages
of this work.  B.~Jo\'o would also like to thank A.~D.~Kennedy for
many useful discussions on the subject.

\end{document}